\documentclass[preprint,nocopyrightspace,10pt]{sigplanconf}

\usepackage{amsmath}
\usepackage{graphicx}
\usepackage{xcolor}
\usepackage{listings}

\lstdefinestyle{cstyle}{frame=tb,
escapechar={@},
  language=c,
  aboveskip=3mm,
  belowskip=3mm,
  showstringspaces=false,
  columns=flexible,
  basicstyle={\small\ttfamily},
  numbers=none,
  numberstyle=\tiny\color{gray},
  keywordstyle=\color{blue},
  keywordstyle=[2]\color{dkgreen},
  keywordstyle=[3]\color{magenta},
  commentstyle=\color{gray},
  stringstyle=\color{mauve},
  breaklines=true,
  breakatwhitespace=true,
  tabsize=3,
  keywords=[2]{def,input,output,inout,kernel,adapt,itn,map,foldl,zipWith,scanl,stendil1D,stencil2D,splitEvery,zip,varinfo,mem,alloc,calloc,realloc,free,copy,io},
  keywords=[3]{polca},
  keepspaces=true
}

\lstdefinelanguage
   [x64]{Assembler}     % add a "x64" dialect of Assembler
   [x86masm]{Assembler} % based on the "x86masm" dialect
   {morekeywords={CDQE,CQO,CMPSQ,CMPXCHG16B,JRCXZ,LODSQ,MOVSXD, %
                  POPFQ,PUSHFQ,SCASQ,STOSQ,IRETQ,RDTSCP,SWAPGS, %
                  rax,rdx,rcx,rbx,rsi,rdi,rsp,rbp, %
                  r8,r8d,r8w,r8b,r9,r9d,r9w,r9b,r10,r10d, %
                  r11,r11d,r12,r12d,r13,r13d}} % etc.

\begin{document}

\setlength{\pdfpageheight}{\paperheight}
\setlength{\pdfpagewidth}{\paperwidth}

\conferenceinfo{CONF 'yy}{Month d--d, 20yy, City, ST, Country}
\copyrightyear{20yy}
\copyrightdata{978-1-nnnn-nnnn-n/yy/mm}
\copyrightdoi{nnnnnnn.nnnnnnn}

% Uncomment the publication rights you want to use.
%\publicationrights{transferred}
%\publicationrights{licensed}     % this is the default
%\publicationrights{author-pays}

%\titlebanner{banner above paper title}        % These are ignored unless
%\preprintfooter{short description of paper}   % 'preprint' option specified.

\preprintfooter{PROHA'16, March 12, 2016, Barcelona, Spain}

\title{Optimized Polynomial Evaluation with Semantic Annotations}
%%%\subtitle{aaa}

\authorinfo{Daniel Rubio Bonilla\and Colin W. Glass}
           {HLRS - University of Stuttgart}
           {rubio@hlrs.de / glass@hlrs.de}
\authorinfo{Jan Kuper}
           {University of Twente}
           {j.kuper@utwente.nl}

\maketitle

\begin{abstract}

In this paper we discuss how semantic annotations can be used to introduce mathematical algorithmic information of the underlying imperative code to enable compilers to produce code transformations that will enable better performance. By using this approaches not only good performance is achieved, but also better programmability, maintainability and portability across different hardware architectures. To exemplify this we will use polynomial equations of different degrees.

\end{abstract}


% general terms are not compulsory anymore,
% you may leave them out


\section{Introduction}

Code optimization is the process that tries to improve the code by making it consume less resources such us CPU cycles, memory, or communication in distributed systems. The word optimization comes from the root ``optimal'' (which comes from the Latin word \textit{optimus}), meaning that it cannot be better. But it is very rare that this process can produce really optimal code. The optimized code can in most cases be optimal for a given use in a determined hardware system. One can often reduce the execution time by making it consume more memory, but in systems where the memory space is scare it might be beneficial to code a slower algorithm which reduces memory usage. In summary, in most cases there is ``no size that fits all'' code that executes optimal in all cases. Software developers that write code for generic systems must adjust their code to perform reasonably well in most common situations.

The code optimization process can also take big amount of time, that can be thought as a cost; it is possible that beyond a certain level of optimization it is not cost effective to invest more time in improving the execution performance. Software developer time is not the only price to pay for code optimization as often the process will lead to source code that is obfuscated, more difficult to maintain or modify and reduces the portability across different hardware systems. Another drawback is that it will also reduce the opportunities to reuse certain parts of the code.

%In large scale or embedded systems energy consumption is usually an important factur, and reducing the energy consumed, even when performance is satisfactory, might save costs (bill, or battery size needs).

To deal with all this factors the ideal situation would be when the source code of the application resembles as close as possible the mathematical formulation of the algorithm, leaving the optimization of the code to the compiler. Unfortunately current compilers are not able to make the same level of aggressive transformations on a near to mathematics code because it is not aware of the intention of the computation (i.e. what is the final result that we want to achieved). For this reason the optimizations are limited to a small set of optimizations that guarantee that the computation is correct.

Introducing mathematical information to the compiler can allow it to perform optimizations at the algorithmic level, similarly to what a software developer might do, to better exploit the characteristics of the underlying hardware. After this step the usual set of current optimizations can be applied.

In this paper we will show how the C resembling the formulation of equivalent polynomial functions affects the execution time and we will discuss how the mathematical information can be introduced through semantic annotations into programming models so that it can be exploited by compilers, allowing them to perform this set of algorithmic optimizations.

\section{Polynomial Representations}
In mathematics, a polynomial is an expression consisting of variables and coefficients that involves only the operations of addition, subtraction, multiplication and non-negative integer exponents. Polynomials appear in a wide range of problem complexity of different areas. In mathematics they are used in calculus and numerical analysis to approximate to other functions, or in advanced mathematics they are used to construct polynomial rings and algebraic varieties, central concepts in algebra. 
They are also often used in physics and chemistry to, for example, describe the trajectory of projectiles, express equations such as the ideal gas law or, polynomial integrals (the sums of many polynomials), can be used to express energy, inertia and voltage difference, to name a few applications. Other natural sciences also use polynomial, like the construction of astronomical and meteorological models.
As polynomials are very useful to express curves, thus engineers use them to design roads, bridges, railways lines, and roller coasters. Combinations of polynomial functions can be used in more sophisticated analysis to retrieve more data, for this, they are applied in the field of economics to do cost analysis.

Mathematical equations can often be expressed under different formulations that are functionally equivalent, but that if taken directly into code will be executed in a different way, resulting in a different resource usage of the available hardware. To illustrate this we will present a polynomial equations and transform its mathematical representation in different steps. The resulting different, but equivalent, functions will be \textit{directly} coded in C language and then evaluated against each other.

Let the quartic polynomial function (degree four) be given by:\\

$f_0(x) = A_4 x^4 + A_3 x^3 + A_2 x^2 + A_1 x^1 + A_0$\\

\begin{figure}[ht!]
  \centering
  \includegraphics[scale=0.45]{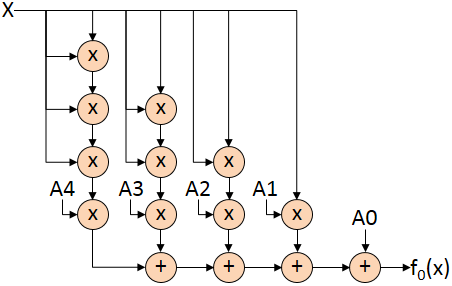}
  \caption{Execution model of $f_0(x)$}
  \label{pic:poly1}
\end{figure}

In Figure~\ref{pic:poly1} the execution structure, \textit{i.e.} data/work flow, of this polynomial function is graphically represented: $x^i$ is represented by a sequence of multiplications with $x$, starting from $x \cdot x$, and the result is multiplied with $a_i$. Then, the results for every $i$ are added.

Listing~\ref{lst:poly1} shows the source code implementation in C that matches the execution model of the equation $f_0(x)$.

\begin{lstlisting}[style=cstyle,caption={$f_0(x)$ C implementation},label={lst:poly1}][ht!]
#define A0 ...
...

int polyCalc(int x) {
  int res;

  res = A4*x*x*x*x + A3*x*x*x + A2*x*x + A1*x + A0;

  return res;
}
\end{lstlisting}

However, this execution model is rather inefficient, as it will be shown in the benchmark section, as the outcomes of all terms $x^i$ are calculated separately. Instead, we may calculate the corresponding values $x_i = x^i$ incrementally:\\

$x_1 = x,  x_2 = x x_1,  x_3 = x x_2,  x_4 = x x_3$\\

\noindent and then define an improved version of the polynomial function:\\

$f_1(x) = A_4 x_4 + A_3 x_3 + A_2 x_2 + A_1 x_1 + A_0$\\

\begin{figure}[ht!]
  \centering
  \includegraphics[scale=0.45]{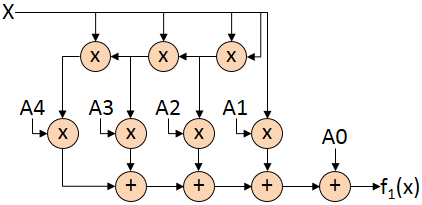}
  \caption{Execution model of $f_1(x)$}
  \label{pic:poly2}
\end{figure}

Figure~\ref{pic:poly2} shows this second execution model where the $x^i$ terms are being reused.

And Listing~\ref{lst:poly2} shows the source code that represents this execution model directly, with the extra variable \texttt{\_x} that stores the incrementally exponentiation.

\begin{lstlisting}[style=cstyle,caption={$f_1(x)$ C implementation},label={lst:poly2}][ht!]
#define A0 ...
...

int polyCalc(int x) {
  int res, _x;

  res = A0;
  res += A1*x;
  _x = x*x;
  res += A2*_x;
  _x *=x;
  res += A3*_x;
  _x *=x;
  res += A4*_x;

  return res;
}
\end{lstlisting}

\begin{figure}[ht!]
  \centering
  \includegraphics[scale=0.45]{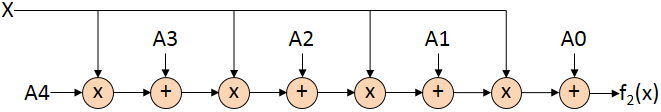}
  \caption{Execution model of $f_2(x)$}
  \label{pic:poly3}
\end{figure}

But it can still done better.
The function can also be rewritten to further reduce the number of multiplications. This is the function $f_2(x)$ that also specifies an execution model, given in Figure~\ref{pic:poly3}. This model is apparently more efficient than the previous ones, in the sense that fewer operations are required, although later we will test this empirically.\\

$f_2(x) = (((A_4 x + A_3)x + A_2)x + A_1)x + A_0$\\

And its execution model is expressed graphically in Figure~\ref{pic:poly3}.

The direct implementation of $f_2(x)$ in C code is shown in Listing~\ref{lst:poly3}.

\begin{lstlisting}[style=cstyle,caption={$f_2(x)$ C implementation},label={lst:poly3}][ht!]
#define A0 ...
...

int polyCalc(int x) {
  int res;

  res = (((A4*x + A3)*x + A2)*x + A1)*x + A0;

  return res;
}
\end{lstlisting}

As the equivalence of $f_0(x)$, $f_1(x)$, and $f_2(x)$ can be proven %TODO: reference?
and the correspondence between the function definitions, the execution model and the derived C code is direct, this also means that the equivalence of the corresponding execution model and C code is guaranteed (as long as we do not care for the numeric differences that can appear due to the lost precision and rounding of floating point variables or cater for possible overflows in integer operations when executing on CPUs).

\section{Code Analysis}
The three mathematical-equivalent formulations have a different number of operations to be performed when directly implemented on C code. The number of operations has a direct impact on the performance on configurable hardware, such us FPGAs~\cite{kuper14a} but the situation might be different on CPUs due to their complex behavior and diversity of architectures. The impact of a particular code is very difficult to predict in the case of superscalar architectures with out-of-order execution. In this chapter we will analyze the theoretical number of computations to be performed for each version and in next chapters we will measure the performance empirically.

The number of operations can be directly counted from the mathematical representations or the C code previously shown for each version. The summary of theoretical operations is summarized in Table~\ref{tbl:compev-05}. As it can be observed the number of additions is constant but the number of multiplications has been significantly reduced with each new representation of the polynomial function.
\begin{table}[ht]
  \begin{center}
    \begin{tabular}{| l | c | c |}
      \hline
       & ADDs & MULs \\ \hline
      $f_0(x)$ & 4 & 10 \\ \hline
      $f_1(x)$ & 4 & 7  \\ \hline
      $f_2(x)$ & 4 & 4 \\ \hline
    \end{tabular}
    \caption{Computational analysis}
    \label{tbl:compev-05}
  \end{center}
\end{table}

When generating the binary code that is later going to be executed on a CPU, compilers do not translate directly the C instructions but perform a set of generic and architecture-specific optimizations that aim to improve the performance over a naive (or direct) code compilation. In this work we have explored the code generated by GCC and LLVM compilers using aggressive optimizations, \texttt{-O3}, and enabling integer operation re-ordering, \texttt{-fstrict-overflow}. 
This last flag allows the compiler to assume strict signed overflow rules. For C and C++ this means that overflow when doing arithmetic with signed numbers is undefined, which means that the compiler may assume that it does not happen. This permits various optimizations, for example, the compiler assumes that an expression like $i + 10 > i$ is always true for signed $i$. This assumption is only valid if signed overflow is undefined, as the expression is false if $i + 10$ overflows when using twos complement arithmetic. For our case it means that the compiler can change, re-order or merge the operations with signed integers~\cite{GCCOO492}.

%% This is for FP variables
%This last compiler flag is necessary because otherwise the compiler would not perform optimizations around floating point variables to preserve numerical equivalence, but we are not interested in this. The different versions are mathematical equivalent but are not numerical equivalent, it is are property that we have already lost and that we are not interested in keeping for the aim of this work.

\begin{table}[ht]
  \begin{center}
    \begin{tabular}{| c | c | c | c | c | c | c |}
      \hline
       & \multicolumn{3}{ |c| }{GCC 4.9.2} & \multicolumn{3}{ |c| }{LLVM 3.6} \\ \hline
      Version & $f_0(x)$ & $f_1(x)$ & $f_2(x)$ & $f_0(x)$ & $f_1(x)$ & $f_2(x)$\\ \hline
      ADD & 4 & 4 & 4 & 4 & 4 & 4  \\ \hline
      MUL & 10 & 7 & 4 & 6 & 6 & 4  \\ \hline
    \end{tabular}
    \caption{Compiler computational analysis with flags \texttt{-O3 -fstrict-overflow}}
    \label{tbl:compev-05cc}
  \end{center}
\end{table}

Table~\ref{tbl:compev-05cc} reflects the number of multiplications and additions that were found in the binary code, after examining the assembler code, generated by GCC and LLVM for different versions of the C code. If we compare this data with Table~\ref{tbl:compev-05}, which contains the number of operations explicitly written in the C code, we can observe that GCC has preserved all the operations while LLVM has been reduced the amount of multiplications for $f_0(x)$ and $f_1(x)$.

For better understanding of the behavior of the compilers we decided to try the same approaches with larger polynomials, in this case of degree 9. For this we can define the polynomial function\\

$g_0(x) = A_9•x^9 + ... + A_1•x^1 + A_0$\\

\noindent and we have produce two equivalent representations,\\

$g_1(x) = A_9 x_9 + ... + A_1 x_1 + A_0$\\

\noindent where\\

$x_1 = x,  x_2 = x x_1,  ...,  x_9 = x x_8$\\

\noindent and finally the version with least operations\\

$g_2(x) = ((A_9 x + A_8)x + ... + A_1)x + A_0$\\

We are not listing the C code of $g_0(x)$, $g_1(x)$ and $g_2(x)$ functions as these are mechanical extensions over the previously shown for $f_0(x)$, $f_1(x)$ and $f_2(x)$ shown in Listings~\ref{lst:poly1},~\ref{lst:poly2} and~\ref{lst:poly3} respectively. The total amount of operations found in these expressions and their direct C implementation is given in Table\ref{tbl:compev-10}, where we can observe that the additions are kept equal but the multiplications can be greatly reduced.

\begin{table}[ht]
  \begin{center}
    \begin{tabular}{| l | c | c |}
      \hline
       & ADDs & MULs \\ \hline
      $g_0(x)$ & 9 & 45 \\ \hline
      $g_1(x)$ & 9 & 17  \\ \hline
      $g_2(x)$ & 9 & 9 \\ \hline
    \end{tabular}
    \caption{Computational analysis}
    \label{tbl:compev-10}
  \end{center}
\end{table}

As with the 4\textsuperscript{th} degree polynomial function we analyzed the binary generated by the compilers. Similarly to the previous case GCC made a direct implementation of the C code but LLVM reduced the number of multiplications, interestingly it was able to produce a binary code with less multiplications for $g_0(x)$ but for $g_1(x)$. But for both compilers the code with less operations comes from the code of $g_2(x)$.

\begin{table}[ht]
  \begin{center}
    \begin{tabular}{| c | c | c | c | c | c | c |}
      \hline
       & \multicolumn{3}{ |c| }{GCC 4.9.2} & \multicolumn{3}{ |c| }{LLVM 3.6} \\ \hline
      Version & $g_0(x)$ & $g_1(x)$ & $g_2(x)$ & $g_0(x)$ & $g_1(x)$ & $g_2(x)$\\ \hline
      ADD & 9 & 9 & 9 & 9 & 9 & 9  \\ \hline
      MUL & 45 & 17 & 9 & 14 & 16 & 9  \\ \hline
    \end{tabular}
    \caption{Compiler computational analysis with flags \texttt{-O3 -fstrict-overflow}}
    \label{tbl:compev-10cc}
  \end{center}
\end{table}

\subsection{LLVM Optimizations}
According to Tables~\ref{tbl:compev-05} and~\ref{tbl:compev-10} the LLVM compiler is able to reduce the number of operations on the original implementation of both polynomial functions $f_0(x)$ and $g_0(x)$ as well as of the modified versions $f_1(x)$ and $g_1(x)$. In this chapter we analyze the generated code by LLVM and compare it to the generated for the code $f_2(x)$ and $g_2(x)$.

\subsubsection{Polynomial Function Degree 4}

The assembler generated by LLVM for the code of the implementation of $f_0(x)$ is shown in Listing~\ref{lst:asmllvm05}.

\begin{lstlisting}[language={[x64]Assembler},caption={LLVM ASM for $f_0(x)$},label={lst:asmllvm05}][ht!]
polyCalc:
	...
	imull	%edi, %r8d         ; i1
	movl	%edi, %eax         ; i2
	imull	%eax, %eax         ; i3
	imull	%eax, %eax         ; i4
	imull	%r9d, %eax         ; i5
	addl	%ecx, %r8d         ; i6
	imull	%edi, %r8d         ; i7
	addl 	%edx, %r8d         ; i8
	imull	%edi, %r8d         ; i9
	leal	(%rax,%rsi), %eax  ; i10
	addl	%r8d, %eax         ; i11
        ...
	retq
\end{lstlisting}

\begin{figure}[ht!]
  \centering
  \includegraphics[scale=0.45]{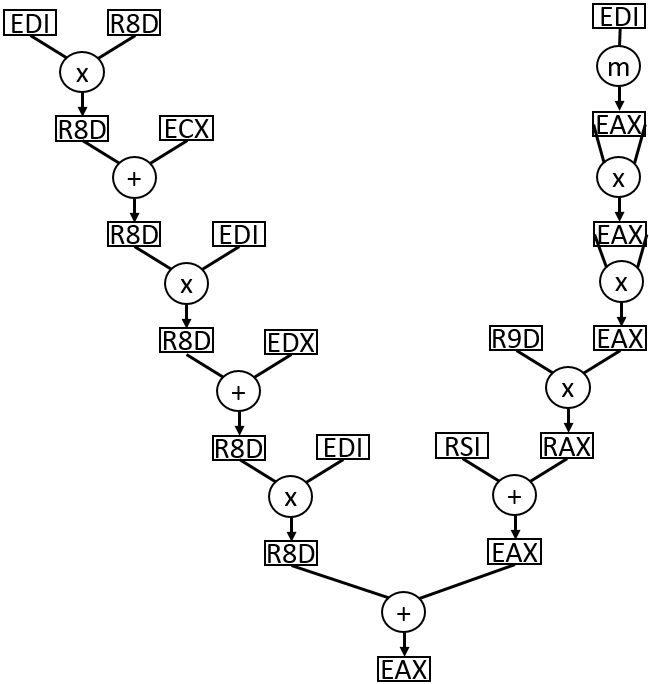}
  \caption{Data and operation graph for LLVM's $f_0(x)$ binary}
  \label{pic:asmllvm}
\end{figure}

The data dependency between registers and operations have been graphically depicted in Figure~\ref{pic:asmllvm}.

From the assembler code and the dependency diagram we can observe that there are two independent execution branches that merge in the last operation. The \textit{left} branch calculates $A_3 x^3 + A_2 x^2 + A_1 x^1$ and the \textit{right} branch calculates $A_4 x^4 + A_0$, adding them together in the last instruction. The assembler lines \texttt{i1} and \texttt{i6} to \texttt{i9} correspond to the left branch while the lines \texttt{i2} to \texttt{i5} and \texttt{i10} correspond to the right branch of the graph.

It is interesting to notice that the maximum path length of this assembler code is 6 instructions, while $f_2(x)$ shows a continuous hard dependency on the previous calculated results, making it a single branch of 8 instructions length. At this point one could inevitable think that LLVM could be trying to exploit superscalar execution models that could take advantage of shorter path even when the total amount of instructions is larger.

\subsubsection{Polynomial Function Degree 10}
The assembler generated by LLVM for the code of the implementation of $g_0(x)$ is shown in Listing~\ref{lst:asmllvm10}.

\begin{lstlisting}[language={[x64]Assembler},caption={LLVM ASM for $g_0(x)$},label={lst:asmllvm10}][ht!]
polyCalc:
	imull	%edi, %r8d
	movl	%edi, %eax
	imull	%eax, %eax
	movl	%eax, %r11d
	imull	%r11d, %r11d
	imull	%r11d, %r9d
	movl	24(%rsp), %r10d
	imull	%r11d, %r10d
	imull	%edi, %eax
	imull	%eax, %eax
	movl	32(%rsp), %ebx
	imull	%eax, %ebx
	imull	40(%rsp), %eax
	imull	%r11d, %r11d
	movl	48(%rsp), %ebp
	imull	%r11d, %ebp
	imull	56(%rsp), %r11d
	addl	%ecx, %r8d
	imull	%edi, %r8d
	addl	%edx, %r10d
	addl	%eax, %r10d
	addl	%r11d, %r10d
	addl	%r8d, %r10d
	imull	%edi, %r10d
	leal	(%r9,%rsi), %eax
	addl	%ebx, %eax
	addl	%ebp, %eax
	addl	%r10d, %eax
	...
	retq
\end{lstlisting}

As happened with the degree 4 polynomial function, the binary generated by LLVM for this case, contains more operations than the compilation of $g_2(x)$ but in different data-independent branches instead of a single branch. The $g_2(x)$ branch has a total of 18 operations while the $g_0(x)$ LLVM optimized code longest branch has 12 operations. In the next section of this work we will evaluate the performance of the different generated binaries.

\section{Benchmarks}

Although the analysis of the theoretical computational of the different equivalent polynomial representations and the binary code generated by the different compilers is interesting the most important is to actually run the different binaries and measure its performance. All the benchmarks were run on a Intel i7-4770k CPU, based on the Haswell architecture, with energy saving (sleeping states and throttling disabled) and core and uncore parts locked at 2 GHz and 2x8 GiB of DDR3 RAM at 1600 MHz (CL9). As we have the clock of the CPU completely locked we have just measured the cycles that were needed to execute each polynomial function (the more cycles needed the slower the code was).

The cycles were obtained by reading the Time Stamp Counter (\texttt{TSC}). It is a 64-bit register present in most modern x86 processors that counts the number of cycles since reset and can be read using the instruction \texttt{RDTSC} that returns the \texttt{TSC} value in \texttt{EDX:EAX} registers (or \texttt{RDTSCP} that forces every preceding instruction to be completed in out-of-order CPUs). In Haswell based CPUs the
\texttt{TSC} register increments at a constant rate set by the maximum resolved frequency at which the processor is booted, 2 GHz in our case, and is synchronized across all cores of the CPU (in older CPUs it was not incrementing constantly but varying with the core's actual frequency and values could be different on different cores).

Table~\ref{tbl:benchmark1} shows the cycles needed to resolve 128 polynomial functions of degree 4 when they were written resembling the formulation of $f_0(x)$, $f_1(x)$ or $f_2(x)$ under LLVM and GCC compilers. Note that the results were obtained in a superscalar CPU, it can execute many instructions in a single cycle, thus the amount of cycles used can be lower than the total amount of instructions.

\begin{table}[ht]
  \begin{center}
    \begin{tabular}{| c | c | c | c | c | c | c |}
      \hline
       & \multicolumn{3}{ |c| }{GCC 4.9.2} & \multicolumn{3}{ |c| }{LLVM 3.6} \\ \hline
      Version & $f_0(x)$ & $f_1(x)$ & $f_2(x)$ & $f_0(x)$ & $f_1(x)$ & $f_2(x)$\\ \hline
      Cycles & 1364 & 1000 & \textbf{868} & 976 & 960 & \textbf{852} \\ \hline
    \end{tabular}
    \caption{Cycles needed to resolve 128 polynomal functions of degree 4 -- less cycles is better}
    \label{tbl:benchmark1}
  \end{center}
\end{table}

The results show that the code that resembles $f_2(x)$ formulation of the polynomial function performs the best in both compiler with LLVM being faster by a~1\%. LLVM optimizations on $f_0(x)$ make the code perform much better that GCC's versions, which did not reduce the number of operations, making it almost as performant as LLVM's version of $f_1(x)$. This answers the question that was arrisen in the previous section. In this case, for a Haswell CPU, having two shorter execution data-independent branches was not faster than the single and larger fully-dependent branch of $f_2(x)$. GCC produced code that performed proportionally inverse to the amount of arithmetic instructions, being always slower than LLVM (with a small margin for $f_1(x)$ and in particular for $f_2(x)$).

Similarly, Table~\ref{tbl:benchmark2} shows the cycles needed to resolve 128 polynomial functions of degree 9 when they were written resembling the formulation of $g_0(x)$, $g_1(x)$ or $g_2(x)$ under LLVM and GCC compilers.

\begin{table}[ht]
  \begin{center}
    \begin{tabular}{| c | c | c | c | c | c | c |}
      \hline
       & \multicolumn{3}{ |c| }{GCC 4.9.2} & \multicolumn{3}{ |c| }{LLVM 3.6} \\ \hline
      Version & $g_0(x)$ & $g_1(x)$ & $g_2(x)$ & $g_0(x)$ & $g_1(x)$ & $g_2(x)$\\ \hline
      Cycles & 5804 & 2348 & \textbf{1964} & 2012 & 2184 & \textbf{1908} \\ \hline
    \end{tabular}
    \caption{Cycles needed to resolve 128 polynomal functions of degree 9 -- less cycles is better}
    \label{tbl:benchmark2}
  \end{center}
\end{table}

In this case the $g_2(x)$ version of the code is the fastest for both compilers, being LLVM almost a~3\% faster than GCC. The binary generated by GCC was faster the further the number of arithmetic operations were reduced in the code. This is not the same case for LLVM as $g_0(x)$ performs better than $g_1(x)$. Even if the C implementation of $g_0(x)$ has more operations than $g_1(x)$, LLVM is able to optimize better $g_0(x)$ producing a binary with slightly less arithmetic instrucions, thus performing better. As with the degree 4 polynomial function, for LLVM, $g_2(x)$ performs better than the optimized $g_0(x)$. In Haswell based CPU, the multiple-independent shorter branched $g_0(x)$ binary does not perform as good as the single fully-dependnent branch code of $g_2(x)$.

\section{Semantic Annotations}
The code optimizations needed to move from $f_0(x)$ to $f_1(x)$ and finally to $f_2(x)$ are not possible in current compilers because current programming models do not convey enough information to enable such transformations without the risk of violating correctness of behaviour.
We can say that compilers are missing information about the intention (in terms of ``semantics'') of the programmer and the structure and principle behaviour of the program.

For these reasons, projects such as POLCA are currently researching in providing programming models that convey sufficient information to meet the following goals~\cite{POLCA13}:
\begin{itemize}
\item provide structural (dependencies and operational behaviour) and mathematical information,
\item enable transformation of the source code,
\item allow the toolchain to assess the ``appropriateness'' of the algorithm and transformations for specific hardware platforms
\item and maintain programmability and correctness.
\end{itemize}

To enable transformations exposed in this work we need to provide enough information to the compiler so that it ``understand'' the mathematical properties of the code running, by using semantic annotations, and strongly binding them to the C code so that the later can be correctly manipulated, reformulating the original implemented algorithm and code for an equivalent but with better performance.

A proposal of the annotations applied to the code of Listing~\ref{lst:poly1} is shown in Listing~\ref{lst:acode}. First the \texttt{ring\_prop} (Ring Properties) pragma annotation declares that, for the function \texttt{PolyCalc} that we assume a number of algebraic laws to hold: associativity of $+$ and $*$, that $0$ and $1$ are neutral elements for $+$ and $*$, that $*$ distributes over $+$ for \texttt{int} data types. 
A mathematical ring is one of the fundamental algebraic structures used in abstract algebra. It consists of a set equipped with two binary operations that generalize the arithmetic operations of addition and multiplication. Through this generalization, theorems from arithmetic are extended to non-numerical objects such as polynomials, series, matrices and functions. A ring is an abelian group with a second binary operation that is associative, is distributive over the abelian group operation, and has an identity element. By extension from the integers, the abelian group operation is called addition and the second binary operation is called multiplication~\cite{Noether21}\cite{Fraenkel14}.

In summary, a \textit{ring} is a set $R$ equipped with binary operations $+$ and $\cdot$ satisfying the following three sets of axioms, called the ring axioms~\cite{boubaki70}~\cite{maclane67}~\cite{lang02}:
\begin{enumerate}
\item $R$ is an abelian group under addition, meaning that:
\begin{itemize}
\item $(a + b) + c = a + (b + c)$ for all $a, b, c$ in $R$ ($+$ is associative).
\item $a + b = b + a$ for all $a, b$ in $R$ ($+$ is commutative).
\item There is an element $0$ in $R$ such that $a + 0 = a$ for all $a$ in $R$ ($0$ is the additive identity).
\item For each a in $R$ there exists $−a$ in $R$ such that $a + (−a) = 0$ ($−a$ is the additive inverse of $a$).
\end{itemize}
\item $R$ is a monoid under multiplication, meaning that:
\begin{itemize}
\item $(a \cdot b) \cdot c = a \cdot (b \cdot c)$ for all $a, b, c$ in $R$ ($\cdot$ is associative).
\item There is an element $1$ in $R$ such that $a \cdot 1 = a and 1 \cdot a = a$ for all $a$ in $R$ ($1$ is the multiplicative identity).
\end{itemize}
\item Multiplication is distributive with respect to addition:
\begin{itemize}
\item $a \cdot (b + c) = (a \cdot b) + (a \cdot c)$ for all $a, b, c$ in $R$ (left distributivity).
\item $(b + c) \cdot a = (b \cdot a) + (c \cdot a)$ for all $a, b, c$ in $R$ (right distributivity).
\end{itemize}
\end{enumerate}

By stating that these mathematical properties apply we enable the compiler to be able to manipulate the code instructions accordingly without taking into consideration possible side effects that could make the code not valid under certain conditions (\textit{e.g.} integer overflow).

Then the second pragma annotation, \texttt{math\_exp} (Mathematical Expression), states the mathematical computation that is performed in the \texttt{PolyCalc} function to which it applies. In it the terms that appear share the name with the constants are variables used in the code, being the evaluation of the expression the value of the function itself, \textit{i.e.} its return value.

\begin{lstlisting}[style=cstyle,caption={$f_0(x)$ implementation with semantic annotations},label={lst:acode}][ht!]
#define A0 ...
...

#pragma ring_prop (+, 0, -, *, 1) int
#pragma math_exp (A0 + A1*x + A2*x^2 + A3*x^3 + A4*x^4)
int polyCalc(int x) {
  int res;

  res = A4*x*x*x*x + A3*x*x*x + A2*x*x + A1*x + A0;

  return res;
}
\end{lstlisting}

By capturing this information, the mathematical expression and the operators properties, and making the compiler aware of it, the compiler may decide to transform the body of the function, modifying the original code resembling the expression $f_0(x)$ into code that resembles $f_1(x)$ or $f_2(x)$.

\section{Conclusions and Future Work}

In this paper we have shown, by using polynomial equations, that mathematical expressions may have different but equivalent formulations and that we can write code that resembles each of these formulations. The different versions of the code will perform different depending on the compiler used to generate the binary code and the hardware that will execute it.
We have also discussed that it is possible to manually search for a good combination of specific and optimized code, for an specific algorithm, and the hardware where it is going to run. But if later the hardware is changed the performance of the code might be worse that the original not-optimize code due to specific hardware optimizations for the old platform that now ruin performance. For example it has been shown that the $g_1(x)$ code looks more optimal (and complex) from a theoretical point of view than the original $g_0(x)$, and it behaves better when compiled with GCC, but the performance decreases if compiled with LLVM.

To address these issues we have proposed the usage of semantic annotations that introduce mathematical information with the aim to allow the compiler to produce optimize code beyond current capabilities from a generic clean code. This will allow to write programs with less bugs, easier to maintain and easier to extend, while at the same time increasing portability. These annotations have to be further tested and extended to be able to address a wider set of mathematical expressions and enable their usability in future compilers.

Regarding the polynomial functions, we will extend the analysis of the performance effects, in terms of time to completion and energy used, of using floating point data types and not only of integers. These effects should also be measured in other architecture families, such as ARM. It would also be interesting to determine the effects of even larger polynomial degrees and introduce further manually optimize code with independent executable branches trying to exploit superscalar architectures.

%% \appendix
%% \section{Appendix Title}

%% This is the text of the appendix, if you need one.

\acks

This project has received funding from the European Union's Seventh Framework Programme under the POLCA project, Grant number 610686.

% We recommend abbrvnat bibliography style.

\bibliographystyle{abbrvnat}

% The bibliography should be embedded for final submission.

\end{document}